\documentclass[twoside]{article}
\newcommand{\figurecaptionfont}{\bfseries}
\newcommand{\tablecaptionfont}{\bfseries}
\newbox\tempbox
\usepackage{qic}
\usepackage{epsfig}
\usepackage{natbib}
\usepackage{ragged2e}
\usepackage{xcolor}
\usepackage{amsmath}

\usepackage{amsfonts}
\usepackage{graphicx}
\usepackage[margin=1in]{geometry}

\begin{document}

% Setting text height for subsequent pages
\setlength{\textheight}{8.0in}

% Defining running head
\runninghead{Dynamical Evolution of Quantum Correlations and Decoherence in Coupled Oscillators Interacting with a Thermal Reservoir}{S. Mehrabankar et al.}

% Setting page style and counter
\thispagestyle{empty}
\setcounter{page}{1}

% Adding title and author information
\begin{center}
    \vspace*{0.88in}
    \textbf{Dynamical Evolution of Quantum Correlations and Decoherence\\ in Coupled Oscillators Interacting with a Thermal Reservoir} \\
    \vspace*{0.1in}
    %\footnotesize Updated: \today \\
    \vspace*{0.225in}
    \footnotesize Somayeh Mehrabankar \\
    \footnotesize \textit{Queensland Quantum and Advanced Technologies Institute, Griffith University, Yuggera Country, Brisbane, QLD 4111, Australia} \\
    \vspace*{0.225in}
    \footnotesize Farkhondeh Abbasnezhad \\
    \footnotesize \textit{Department of Physics, Faculty of Science, Shahid Chamran University of Ahvaz, Ahvaz, Iran} \\
    \vspace*{0.225in}
    \footnotesize Davood Afshar\\
    \footnotesize \textit{Department of Physics, Faculty of Science, Shahid Chamran University of Ahvaz, Ahvaz, Iran} \\
    \footnotesize \textit{Center for Research on Laser and Plasma, Shahid Chamran University of Ahvaz, Ahvaz, Iran} \\
    \vspace*{0.225in}
    \footnotesize Aurelian Isar \\
    \footnotesize \textit{Department of Theoretical Physics, National Institute of Physics and Nuclear Engineering, Bucharest-Magurele, Romania}
\end{center}

% Adding abstract
\begin{abstract}
    \justifying
    We investigate the dynamical evolution of quantum discord, entanglement and purity in an open quantum system consisting of two coupled asymmetric harmonic oscillators interacting with a thermal environment. Using the Kossakowski-Lindblad master equation we analyze the time evolution starting with a squeezed vacuum state. In contrast to our previous study on entanglement evolution in asymmetric oscillators ~\cite{afshar_entanglement_2016}, the present work introduces the XY-type position-position coupling ($\nu x_1 x_2$) together with a systematic joint analysis of quantum discord and purity alongside entanglement. We examine the combined effects of the squeezing parameter $r$, asymmetry parameter $\varepsilon$, coupling constant $\nu$, dissipation rate $\lambda$ and temperature $T$. We find that quantum discord and entanglement exhibit, in general, a non-monotonic decrease over time. Increasing temperature consistently accelerates the degradation of both quantum correlations and purity, whereas increasing dissipation accelerates the degradation of quantum correlations but leads to higher steady-state purity. Increasing the squeezing parameter provides a protective effect by enhancing the initial correlations and prolonging the survival time of entanglement, increasing the coupling constant leads to higher quantum correlations. The asymmetric parameter, however, exhibits only a weak influence on the correlation evolution. Our analysis reveals that quantum discord demonstrates a stronger resilience than entanglement, which can present a more complex behaviour, including entanglement sudden death and possible temporary revivals and re-suppressions. These findings provide valuable insights for developing robust quantum information protocols and strategies for preserving quantum correlations in realistic open quantum systems, with potential extensions to non-Markovian regimes and multi-mode architectures.
\end{abstract}

% Adding keywords
\vspace*{10pt}
\textbf{Keywords:} Coupled Oscillators, Gaussian states, Quantum Correlations, Purity, Decoherence, Open Quantum Systems, Markovian master equation, XY coupling
% Starting main content
\section{Introduction}
Continuous variable systems are central to quantum optics and quantum information protocols, playing crucial roles in applications such as quantum teleportation, quantum key distribution, and quantum computation \cite{laurat_entanglement_2005, adesso_quantum_2010, karimipour_entanglement_2002, furusawa_unconditional_1998, grosshans_continuous_2002}. In these systems, Gaussian states stand out due to their theoretical significance and experimental accessibility. They can be readily generated and manipulated using standard optical devices, like beam splitters, squeezers, and phase shifters \cite{milburn_quantum_1988, reck_experimental_1994}. Gaussian states encompass a broad class of quantum states, including coherent, squeezed, and thermal states, which are fundamental to the understanding and implementation of quantum information theory and quantum optics \cite{rodo_operational_2008}. Particularly, Gaussian states serve as essential resources for quantum information protocols and provide a suitable framework for exploring quantum correlations \cite{giorda_gaussian_2010, henderson_classical_2001, adesso_entanglement_2007}.

In a realistic physical context quantum states are inevitably exposed to diffusion and dissipation effects, due to the interaction of their systems with the surrounding environment. This interaction leads to the phenomenon of decoherence, during which the information stored in the quantum state is irreversibly transferred to the environment \cite{caldeira_path_1983}. Decoherence poses a significant challenge in quantum information processing, as it leads in general to the degrading in time and even to the suppression of quantum correlations \cite{breuer_theory_2002, weiss_fluorescence_1999, zurek_decoherence_2003}. To preserve the quantum correlations is crucial for the effectiveness of quantum information protocols, where these correlations must persist longer than the time required to execute such protocols. Therefore, a deep understanding of the mechanisms driving decoherence is vital for developing strategies to mitigate or even stop this process in quantum systems \cite{nielsen_quantum_2000}. Recent research has extensively investigated the decoherence and evolution of quantum correlations in Gaussian states within open quantum systems, highlighting their importance in practical quantum information processing \cite{isar_quantum_2007, isar_entanglement_2009, an_non-markovian_2007, adesso_multipartite_2006, dodd_disentanglement_2004, isar_quantum_2013, isar_entanglement_2014, ban_decoherence_2006}.

Coupled oscillators are exemplary models for a wide range of physical systems, such as particles trapped in quantum dots under magnetic fields \cite{sudiarta_solving_2007} or vibrational modes in molecular systems \cite{henderson_classical_2001}. The coupling between oscillators can explain fundamental interactions, such as the van der Waals forces between atoms and molecules \cite{vasile_nonclassical_2010}. Furthermore, the double helix of DNA can be modeled as a chain of coupled oscillators interacting through dipole-dipole forces, where the resulting entanglement plays a pivotal role in maintaining the stability of the molecule \cite{rieper_entanglement_2010}. The generation of entangled cluster states via van der Waals interactions between neutral atoms in an optical lattice has been a topic of considerable interest \cite{kuznetsova_cluster_2012}. Moreover, the experimental control of the coupling between two oscillators using Coulomb interaction in trapped ions has recently been demonstrated \cite{brown_coupled_2011, harlander_trapped-ion_2011}. The impact of repulsive coupling on the discord dynamics between two oscillators immersed in a thermal environment has also been explored, offering insights into the behavior of such open quantum systems ~\cite{isar_generation_2017}. 

In contrast to our previous study on entanglement evolution in asymmetric oscillators \cite{afshar_entanglement_2016}, the present work introduces the XY-type position-position coupling ($\nu x_1 x_2$) together with a systematic joint analysis of quantum discord and purity alongside entanglement. We examine the combined effects of the squeezing parameter $r$, asymmetry parameter $\varepsilon$, coupling constant $\nu$, dissipation rate $\lambda$ and temperature $T$ on all three quantities simultaneously. This unified treatment reveals the protective role of squeezing and coupling, the counter-intuitive increase of steady-state purity with stronger dissipation, and the greater long-time resilience of discord compared with entanglement. These aspects extend recent investigations on Gaussian-state dynamics and quantum correlations in thermal environments~\cite{dhahri_environment_2025,hsiang_entanglement_2022,makarov_linearly_2025,elqars_gaussian_2024,arisoy_hot_2023,hahto_transfer_2025}. Utilizing the Markovian approximation, we derive the system evolution in terms of the covariance matrix via the Kossakowski-Lindblad equation \cite{caldeira_path_1983, gorini_completely_1976}. We then investigate the purity, entanglement, and discord as functions of the system parameters and environmental influences. A comparative analysis of the entanglement complex behaviour, including entanglement sudden death, and possibly temporary destructions and survivals of entanglement, and of the discord evolution is also presented, providing deeper insights into the temporal behaviour of quantum correlations in the system under scrutiny.

The paper is structured as follows. In Sec. 2 we detail the model of the system and its interaction with the environment. Sec. 3 introduces the measures of entanglement and discord for a two-mode Gaussian system. Sec. 4 presents an in-depth analysis of the time evolution of quantum correlations and purity. Finally, in Sec. 5, we discuss the results and draw conclusions of the study.

\section{Two Coupled Harmonic Oscillators Interacting with a Thermal Environment}
We consider a system consisting of two coupled asymmetric harmonic oscillators interacting with a thermal environment. Such a system could model, for instance, a system consisting of two ions of identical mass $m$ held in two separated trapping potentials with mutual Coulomb interaction. The system is described by the following Hamiltonian \cite{brown_coupled_2011}:
\begin{equation} \label{cov}
H_s = \frac{p_1^2 + p_2^2}{2m} + \frac{m}{2}(\omega_1^2 x_1^2 + \omega_2^2 x_2^2) + \nu x_1 x_2.
\end{equation}
Denoting by $p_1$ and $p_2$ the momentum operators of the ions, the first term on the right hand side is the kinetic energy. The second term contains the trapping potentials, $x_1$ and $x_2$ being the displacements of the ions from the external potential minima, and the third term represents the significant contribution coming from the Coulomb interaction between the two ions \cite{brown_coupled_2011}. The negative and positive values of $\nu$ describe the attraction and repulsive interaction, respectively, and this Hamiltonian is physical if the following relation is satisfied \cite{hamdouni_quantum_2010}:
\begin{equation} \label{cov1}
|\nu| \le m \omega_1 \omega_2.
\end{equation}
We write the corresponding frequencies of the oscillators as follows:
\begin{equation} \label{3_}
\omega_1 = \omega \sqrt{1 + \varepsilon}, \quad \omega_2 = \omega \sqrt{1 - \varepsilon},
\end{equation}
where $\varepsilon$ denotes the asymmetry parameter, with $0 \le \varepsilon < 1$. We note that $\varepsilon = 0$ corresponds to a symmetric Hamiltonian.

We suppose that the considered system is immersed in a thermal environment consisting of an infinite number of independent harmonic oscillators, which is in a Gibbs state described by the following density operator \cite{ollivier_quantum_2001}:
\begin{equation} \label{4_}
\rho_{th} = \frac{e^{-\beta H_E}}{\mathrm{Tr}(e^{-\beta H_E})},
\end{equation}
where $H_E$ is the Hamiltonian of the environment, $\beta = (k_B T)^{-1}$, $k_B$ is Boltzmann constant and $T$ is the temperature of the environment. In the following we shall use the system of units with $\hbar = m = k_B = 1$ \cite{afshar_entanglement_2016}.

In the Markovian approximation the time evolution of the considered open system is given by the Kossakowski-Lindblad master equation \cite{gorini_completely_1976, lindblad_generators_1976}:
\begin{equation} \label{scond}
\frac{d}{dt} \rho_s(t) = -i [H_s, \rho_s(t)] + \frac{1}{2} \sum_l ([V_l \rho_s(t), V_l^\dagger] + [V_l, \rho_s(t) V_l^\dagger]),
\end{equation}
where $\rho_s$ is the density operator of the open system. Here $V_l, V_l^\dagger$ are the Lindblad operators that describe the interaction between the system and the environment \cite{davies_generators_1979}:
\begin{equation} \label{ste}
V_l = \sum_{j=1}^2 (\alpha_j^l x_j + \beta_j^l p_j), \quad V_l^\dagger = \sum_{j=1}^2 (\alpha_j^{l*} x_j + \beta_j^{l*} p_j), \quad l = 1,2,3,4,
\end{equation}
where $\alpha_j^l$ and $\beta_j^l$ are complex coefficients. We consider that the system is in a two-mode Gaussian state, defined by the covariance matrix in the standard form \cite{adesso_quantum_2010}:
\begin{equation} \label{ste1}
\sigma = \begin{pmatrix}
A & C \\
C^{\rm T} & B
\end{pmatrix},
\end{equation}
where $A = \mathrm{diag}(a,a)$ and $B = \mathrm{diag}(b,b)$ are the covariance matrices of the first and the second subsystem, respectively, and $C = \mathrm{diag}(c,d)$ is the correlation matrix between the two subsystems. The whole information about the Gaussian state is contained in the symplectic eigenvalues:
\begin{equation} \label{ste2}
\nu_\pm^2 = \frac{\Delta \pm \sqrt{\Delta^2 - 4 I_4}}{2},
\end{equation}
where
\begin{equation} \label{10_}
\Delta = I_1 + I_2 + 2 I_3.
\end{equation}
Here $I_1, I_2, I_3$ and $I_4$ are the symplectic invariants defined as follow:
\begin{equation} \label{11_}
I_1 = \det A, \quad I_2 = \det B, \quad I_3 = \det C, \quad I_4 = \det \sigma.
\end{equation}

The covariance matrix has to satisfy the Robertson-Schr\"odinger uncertainty relation, which can be expressed by the following necessary and sufficient condition in terms of symplectic eigenvalues \cite{adesso_quantum_2010, adesso_continuous_2014}:
\begin{equation} \label{12_}
\nu_- \ge 1.
\end{equation}
The degree of purity of the states is defined by the following expression:
\begin{equation} \label{13_}
\mu = \frac{1}{\nu_+ \nu_-},
\end{equation}
where $\mu$ takes values in the range $0 \le \mu \le 1$, with $\mu = 1$ and $\mu = 0$ corresponding to pure and completely mixed states, respectively.

For the considered Gaussian system with the Hamiltonian (1), the covariance matrix corresponding to the density operator satisfying the master equation (5) is given by \cite{sandulescu_open_1987}:
\begin{equation} \label{14_}
\sigma(t) = e^{Mt} [\sigma(0) - \sigma(\infty)] (e^{Mt})^{\rm T} + \sigma(\infty),
\end{equation}
where $\sigma(0)$ and $\sigma(\infty)$ are the covariance matrices of the initial state and of the asymptotic state of the system, respectively. The matrix
\begin{equation} \label{15_}
M = \begin{pmatrix}
-\lambda & 1 & 0 & 0 \\
-\omega_1^2 & -\lambda & -\nu & 0 \\
0 & 0 & -\lambda & 1 \\
-\nu & 0 & -\omega_2^2 & -\lambda
\end{pmatrix},
\end{equation}
where $\lambda$ is the dissipation constant, is completely determined by the complex coefficients $\alpha_j^l$ and $\beta_j^l$ and the parameters of the system \cite{hamdouni_quantum_2010}. The asymptotic covariance matrix $\sigma(\infty)$ is given by the equation \cite{sandulescu_open_1987}:
\begin{equation} \label{17_}
M \sigma(\infty) + \sigma(\infty) M^{\rm T} = -2D,
\end{equation}
where $D$ is the diffusion matrix given by \cite{sandulescu_open_1987}:
\begin{equation}
D = \mathrm{diag} \left\{ \frac{\lambda}{\omega_1} \coth \frac{\omega_1}{2 T}, \lambda \omega_1 \coth \frac{\omega_1}{2 T}, \frac{\lambda}{\omega_2} \coth \frac{\omega_2}{2 T}, \lambda \omega_2 \coth \frac{\omega_2}{2 T} \right\}.
\end{equation}

In this paper we consider the two-mode squeezed vacuum state as initial state of the system with the covariance matrix given by:
\begin{equation} \label{20_}
\sigma(0) = \begin{pmatrix}
\cosh 2r & 0 & \sinh 2r & 0 \\
0 & \cosh 2r & 0 & -\sinh 2r \\
\sinh 2r & 0 & \cosh 2r & 0 \\
0 & -\sinh 2r & 0 & \cosh 2r
\end{pmatrix},
\end{equation}
where $r$ is the squeezing parameter.

\section{Measures for Quantum Correlations in Gaussian States}
To quantify the degree of entanglement of bipartite Gaussian states, it is appropriate to use logarithmic negativity \cite{adesso_quantum_2010}:
\begin{equation} \label{20}
E_n = \max \left\{ 0, -\log \sqrt{\frac{I_1 + I_2 - 2 I_3 - \sqrt{(I_1 + I_2 - 2 I_3)^2 - 4 I_4}}{2}} \right\},
\end{equation}
where $E_n$ is positive for entangled and 0 for separable states.

Another type of quantum correlations, quantum discord, was introduced by Olivier and Zurek \cite{ollivier_quantum_2001}. Discord is defined as the difference between two expressions of the mutual information which are identical classically, but differ when they are quantified \cite{madhok_quantum_2013, isar_quantum_2013}. This correlation is non-zero for all Gaussian states, excepting product states \cite{madhok_quantum_2013}. For bipartite Gaussian states an explicit expression was obtained for the Gaussian quantum discord \cite{adesso_quantum_2010}:
\begin{equation} \label{21_}
D = f(\sqrt{I_2}) - f(\nu_-) - f(\nu_+) + f(\sqrt{\zeta}),
\end{equation}
where
\begin{equation} \label{22_}
f(x) = \frac{x + 1}{2} \log \frac{x + 1}{2} - \frac{x - 1}{2} \log \frac{x - 1}{2},
\end{equation}
and
\begin{equation} \label{23_}
\zeta = \begin{cases}
\frac{2 I_3^2 + (I_2 - 1)(I_4 - I_1) + 2 |I_3| \sqrt{I_3^2 + (I_2 - 1)(I_4 - I_1)}}{(I_2 - 1)^2}, & \text{if } (I_4 - I_1 I_2)^2 \le (I_2 + 1) I_3^2 (I_1 + I_4), \\
\frac{I_1 I_2 - I_3^2 + I_4 - \sqrt{I_3^4 + (I_4 - I_1 I_2)^2 - 2 I_3^2 (I_1 I_2 + I_4)}}{2 I_2}, & \text{otherwise.}
\end{cases}
\end{equation}

\section{Temporal Analysis of Quantum Correlations and Purity}
In this Section we analyze the time evolution of quantum entanglement, quantum discord and the purity of a system composed of two coupled oscillators interacting with a thermal environment. Starting from an initially pure state, we investigate how the system and environmental parameters influence the behaviour of quantum correlations.

\begin{figure}[ht]
    \centering
    \includegraphics[width=0.45\textwidth]{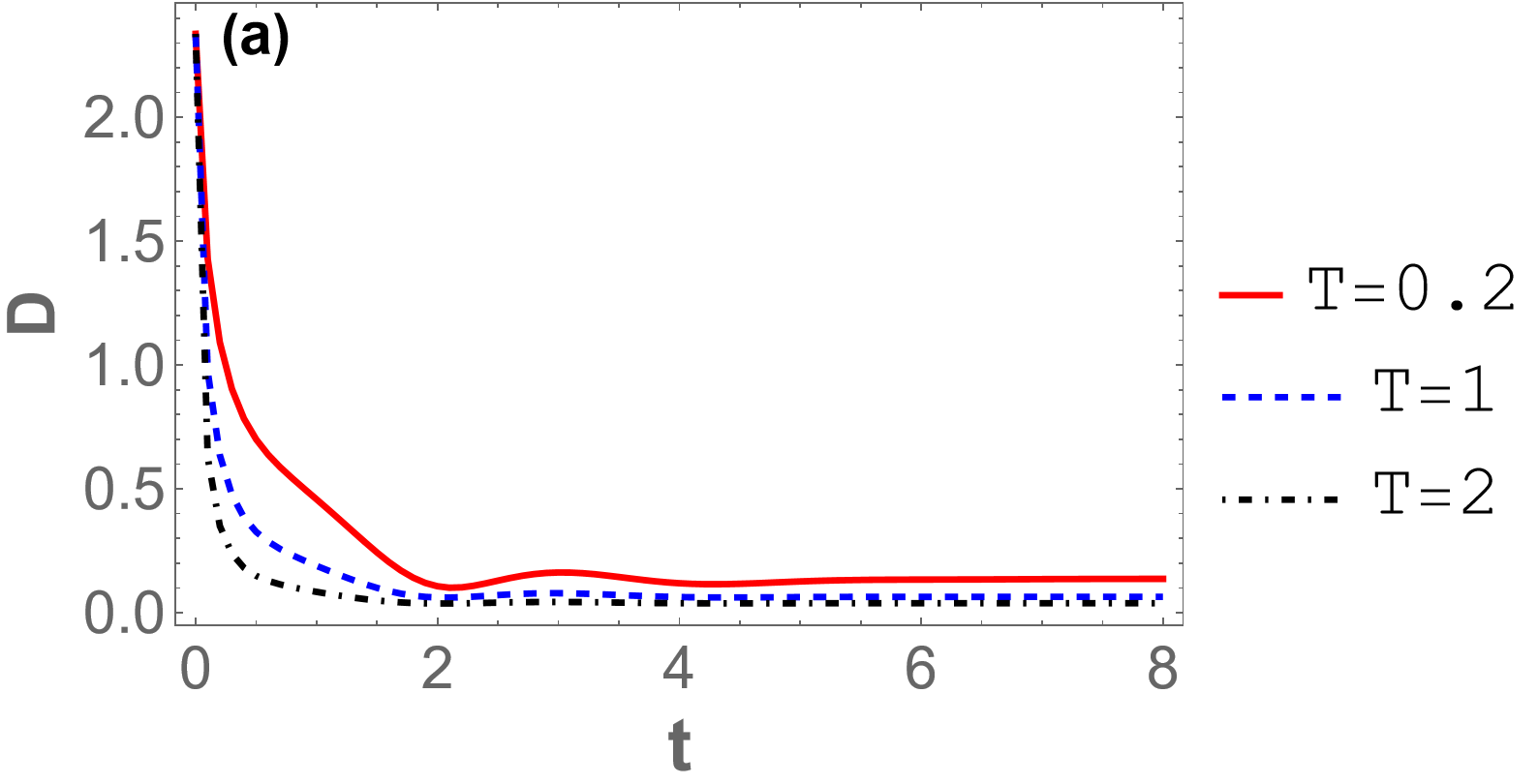}
    \hfill
    \includegraphics[width=0.45\textwidth]{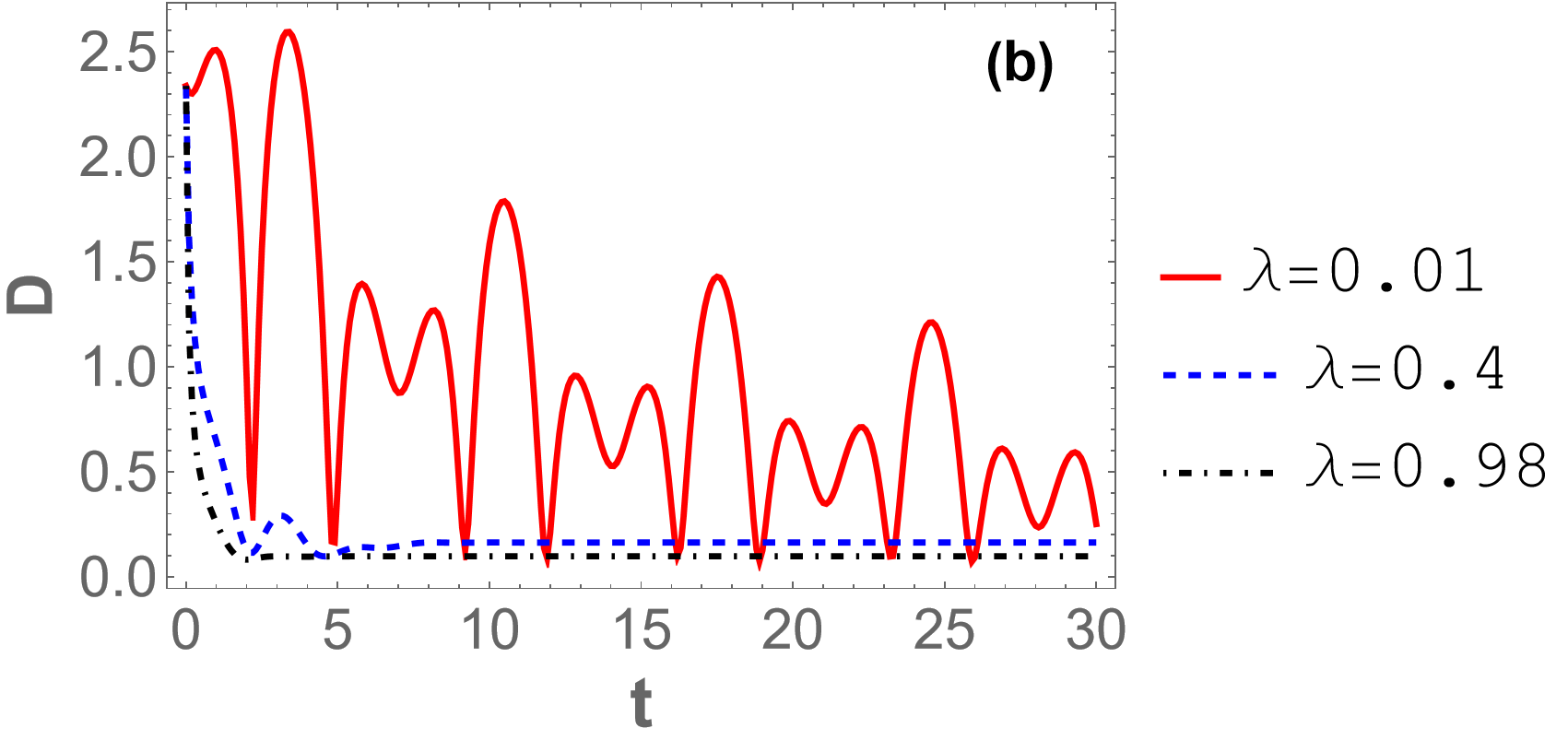}
    \vspace{0.5cm}
    \includegraphics[width=0.45\textwidth]{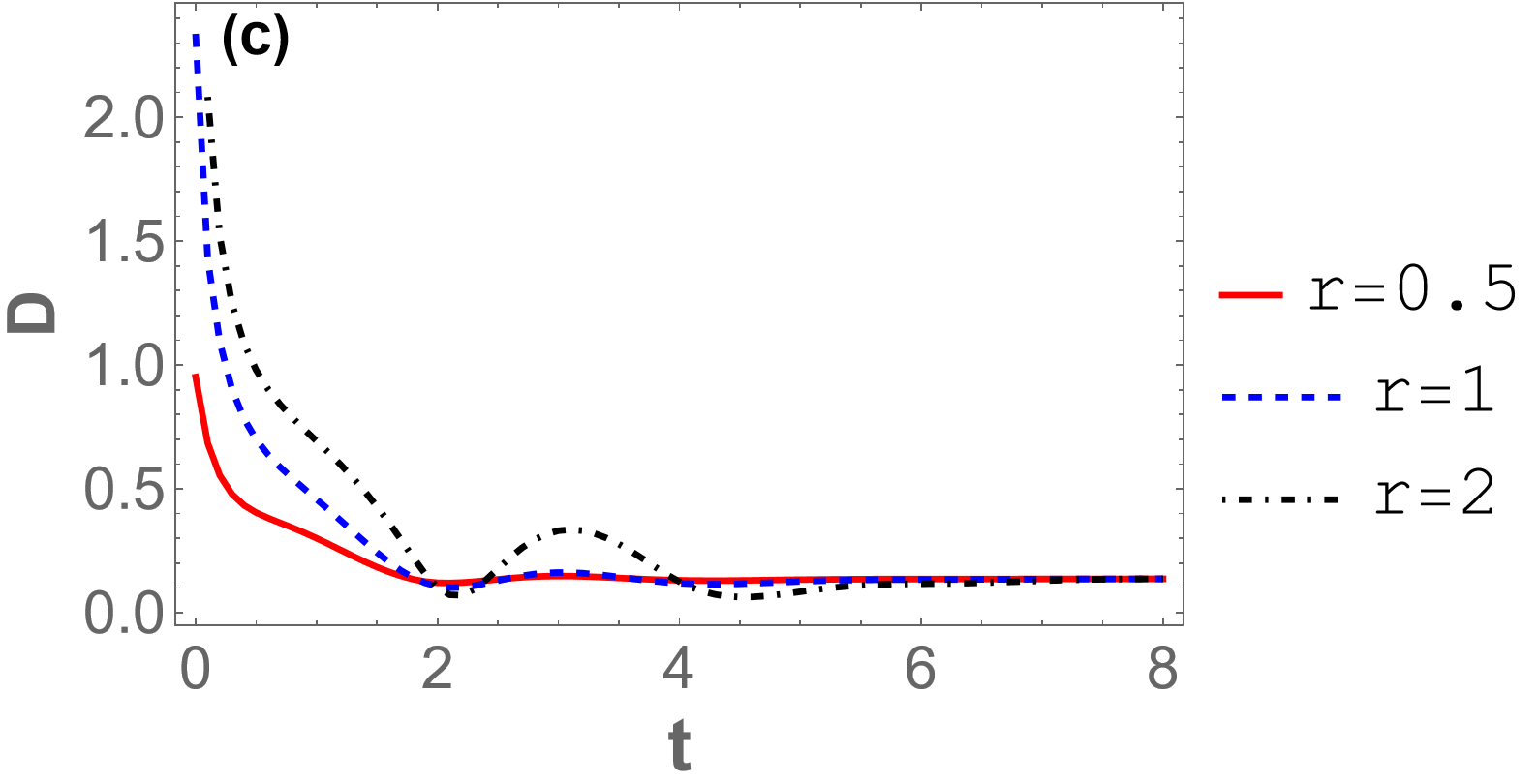}
    \hfill
    \includegraphics[width=0.45\textwidth]{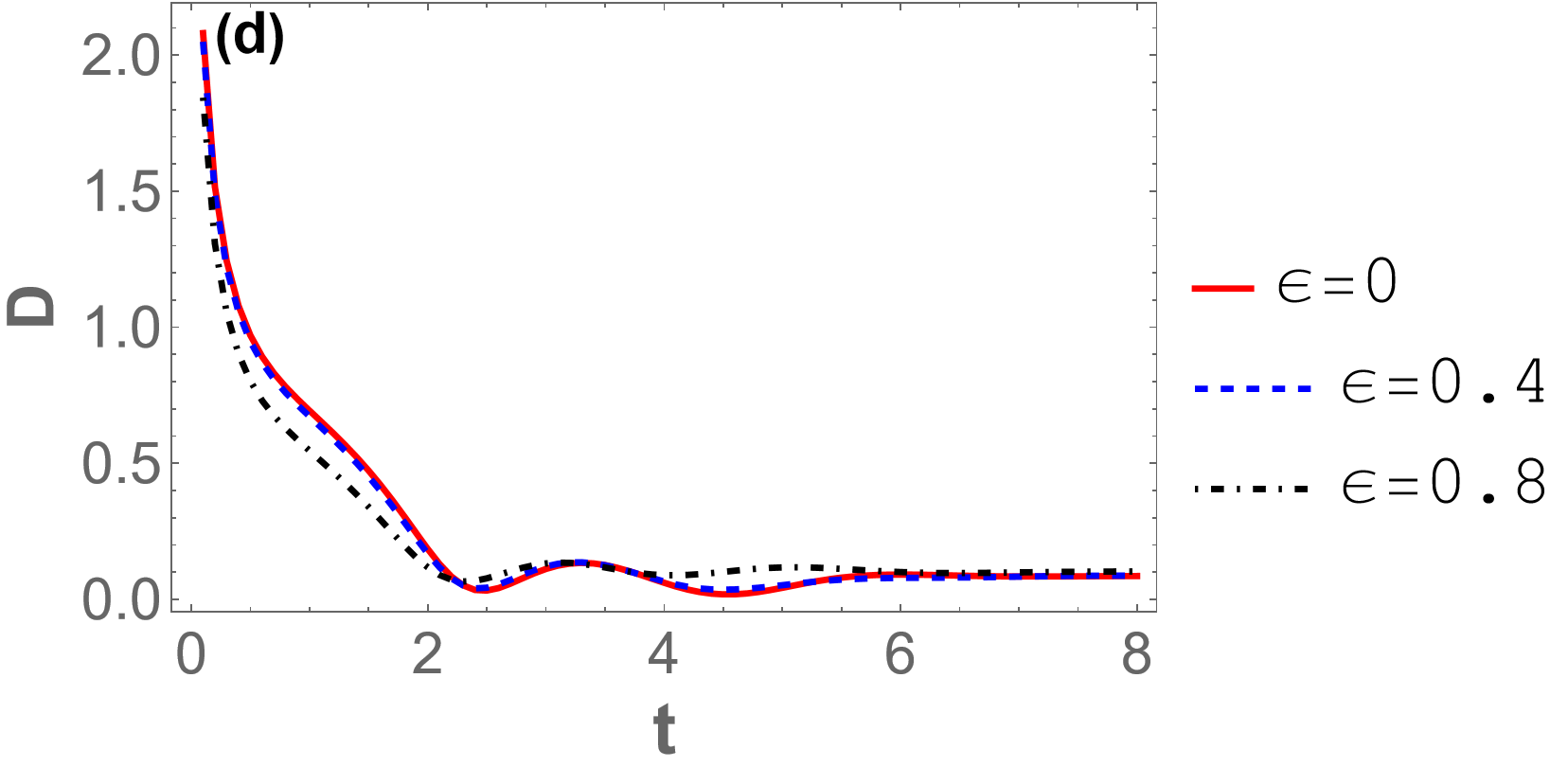}
    \caption{Dynamics of quantum discord for different values of (a) temperature with $\varepsilon=0$, $r=1$, $\lambda=0.6$, $\nu=0.8$, $\omega=1$; (b) dissipation coefficient with $\varepsilon=0$, $r=1$, $T=0.2$, $\nu=0.8$, $\omega=1$; (c) squeezing parameter with $\varepsilon=0$, $T=0.2$, $\lambda=0.6$, $\nu=0.8$, $\omega=1$; (d) asymmetric parameter with $T=0.2$, $r=1$, $\lambda=0.6$, $\nu=0.6$, $\omega=1$.}
    \label{fig:four-figures1}
\end{figure}

As shown in Figs.~1(a)--1(d), quantum discord manifests a general decreasing over time, but the rate and character of the decrease vary. Higher temperatures (Fig.~1(a)) accelerate the decrease of discord and, similarly, an increased dissipation (Fig.~1(b)) leads to a fast decrease of discord. In contrast, larger squeezing parameter (Fig.~1(c)) enhances initial discord, resulting in more pronounced oscillations in time. Asymmetry (Fig.~1(d)) slightly accelerates the initial decrease, but its overall impact is less significant compared to those of temperature and dissipation. Discord keeps a non-zero value all the time, approaching zero asymptotically at infinite time only when the coupling is zero.
\begin{figure}[ht]
    \centering
    \includegraphics[width=0.45\textwidth]{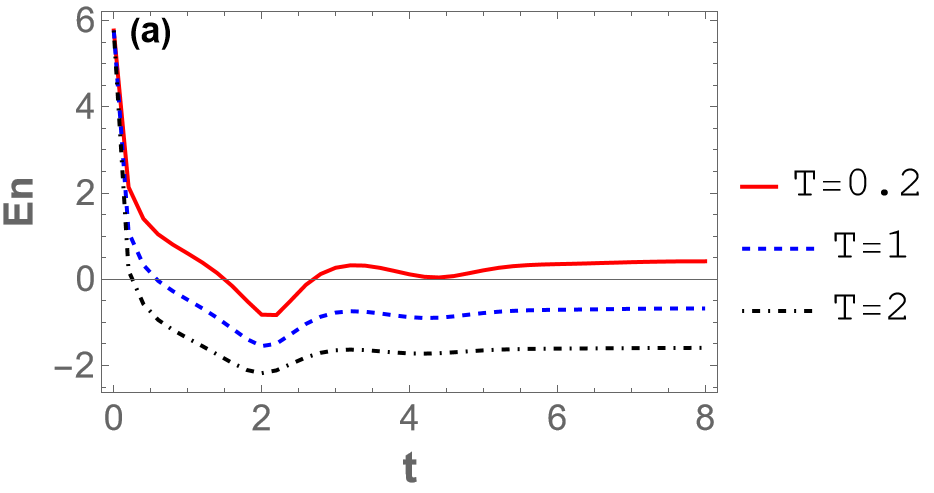}
    \hfill
    \includegraphics[width=0.45\textwidth]{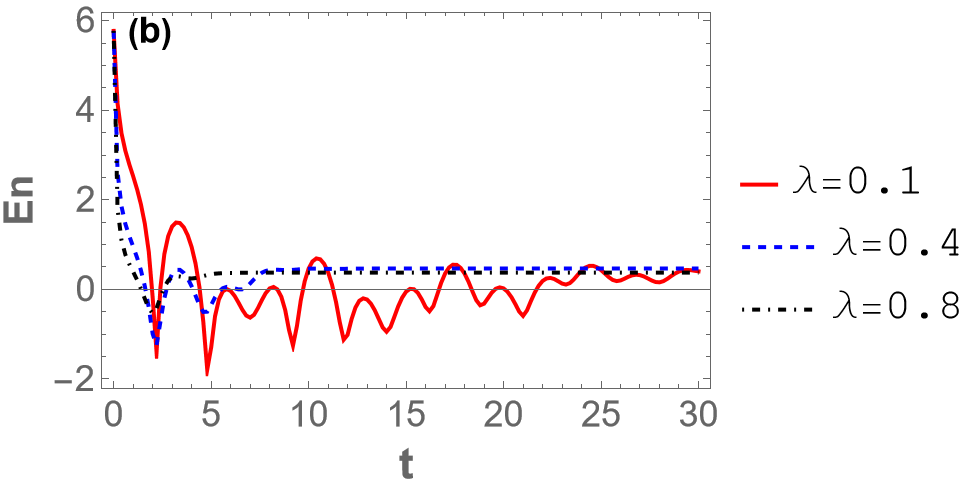}
    \vspace{0.5cm}
    \includegraphics[width=0.45\textwidth]{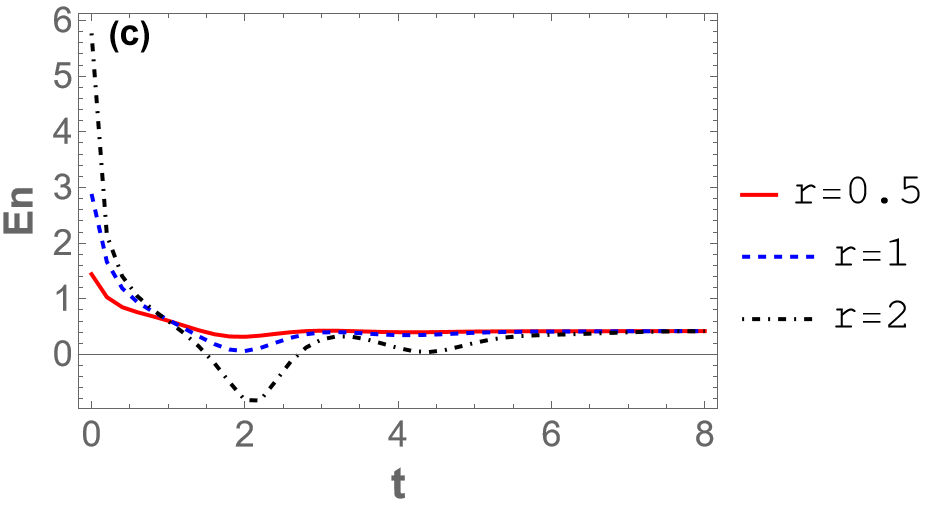}
    \hfill
    \includegraphics[width=0.45\textwidth]{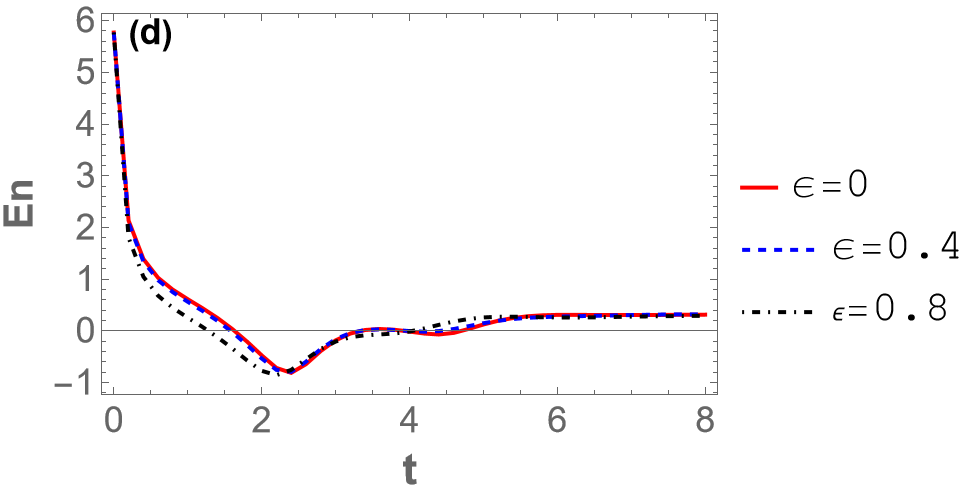}
    \caption{Dynamics of quantum entanglement for different values of (a) temperature with $\varepsilon=0$, $r=2$, $\lambda=0.6$, $\nu=0.8$, $\omega=1$; (b) dissipation coefficient with $\varepsilon=0$, $r=2$, $T=0.2$, $\nu=0.8$, $\omega=1$; (c) squeezing parameter with $\varepsilon=0$, $T=0.2$, $\lambda=0.6$, $\nu=0.8$, $\omega=1$; (d) asymmetric parameter with $T=0.2$, $r=2$, $\lambda=0.6$, $\nu=0.6$, $\omega=1$.}
    %In our system of units, the units of time and temperature are equivalent to $2\,\mathrm{ns}$ and $4\,\mathrm{mK}$, respectively.}
    \label{fig:four-figures2}
\end{figure}

Figures~2(a)--2(d) show that entanglement, measured by logarithmic negativity, exhibits varied behaviour over time. Higher temperatures (Fig.~2(a)) accelerate the decrease of entanglement. Similarly, an increased dissipation (Fig.~2(b)) leads to a fast decrease of entanglement, while larger squeezing parameter (Fig.~2(c)) enhances the entanglement. Depending on the interplay of parameters, entanglement may experience permanent suppression in finite time (``entanglement sudden death''), possibly followed by temporary revivals and suppressions of entanglement, or sustained entanglement. At time infinity, entanglement either persists or has a zero value, this behaviour being determined by the competition between the influences of temperature, dissipation, squeezing and coupling constant. Asymmetry (Fig.~2(d)) has a relatively weak influence on the evolution of the entanglement. 
\begin{figure}[ht]
    \centering
    \includegraphics[width=0.45\textwidth]{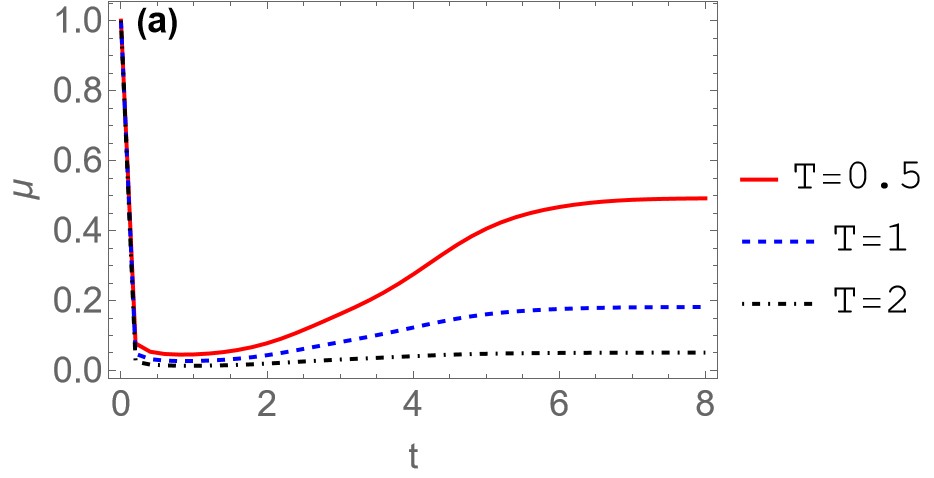}
    \hfill
    \includegraphics[width=0.45\textwidth]{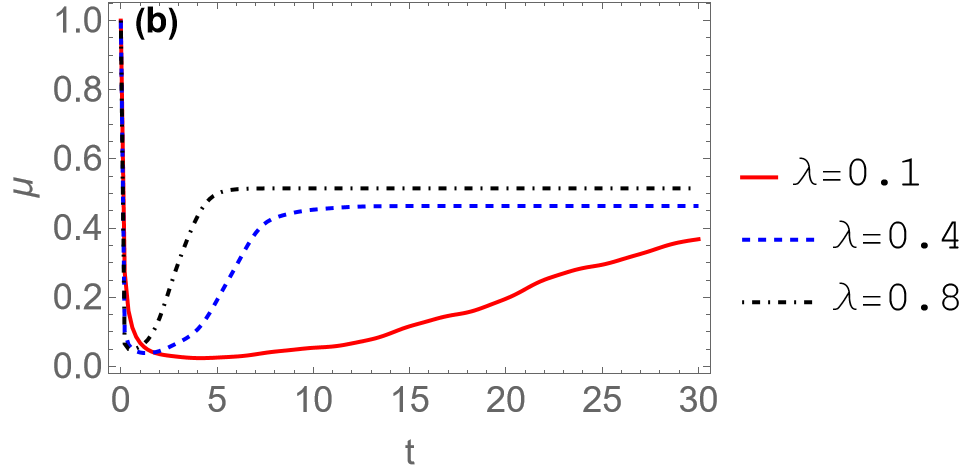}
    \vspace{0.5cm}
    \includegraphics[width=0.45\textwidth]{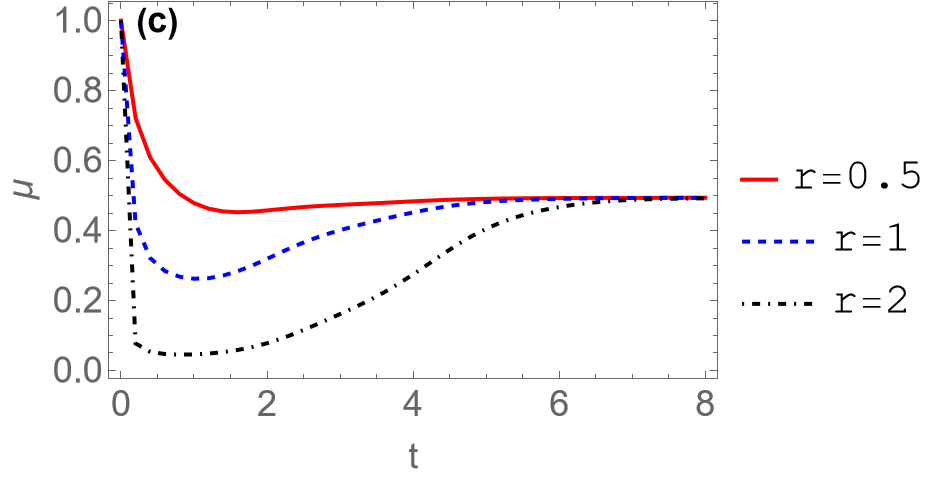}
    \hfill
    \includegraphics[width=0.45\textwidth]{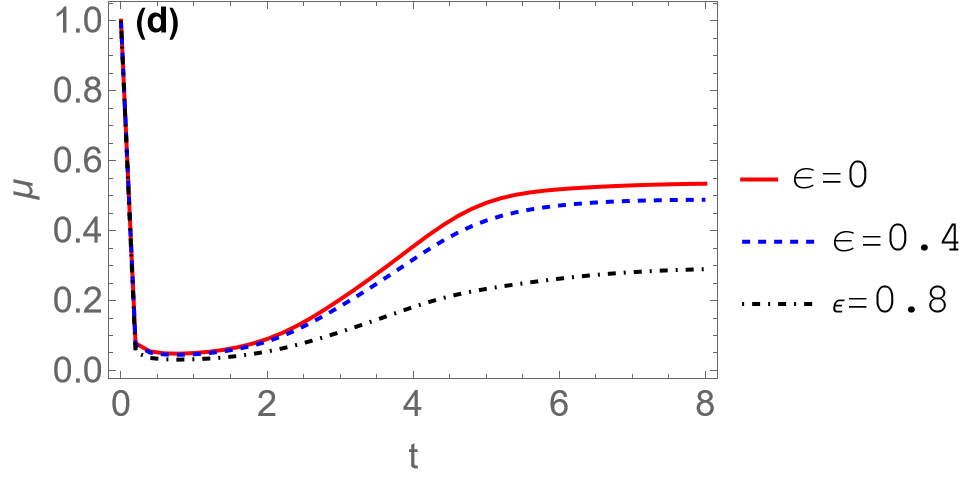}
    \caption{Dynamics of purity for different values of
    (a) temperature with $\varepsilon = 0$, $r = 2$, $\lambda = 0.6$, $\nu = 0.8$, and $\omega = 1$,
    (b) dissipation coefficient with $T = 0.5$, $r = 2$, $\varepsilon = 0$, $\nu = 0.8$, and $\omega = 1$,
    (c) squeezing parameter with $\varepsilon = 0$, $T = 0.5$, $\lambda = 0.6$, $\nu = 0.8$, and $\omega = 1$,
    (d) asymmetric parameter with $\lambda = 0.6$, $r = 2$, $T = 0.5$, $\nu = 0.6$, and $\omega = 1$.}
    \label{fig:four-figures3}
\end{figure}

Figures 3(a)--3(d) depict the evolution of purity, which reflects the degree of mixedness and the decoherence in the system. In all cases, purity decreases initially and then stabilizes. Higher temperatures (Fig. 3(a)) lead to lower steady-state purity, while larger dissipation (Fig. 3(b)) leads to higher steady-state purity, reflecting the complex interplay between dissipation and coupling in the oscillator system. Increasing the squeezing parameter (Fig. 3(c)) also accelerates decoherence, as it amplifies the system sensitivity to noise. At the same time, the asymmetric parameter (Fig. 3(d)) has a relatively weak effect on the purity.

\begin{figure}[ht]
    \centering
    \includegraphics[width=0.45\textwidth]{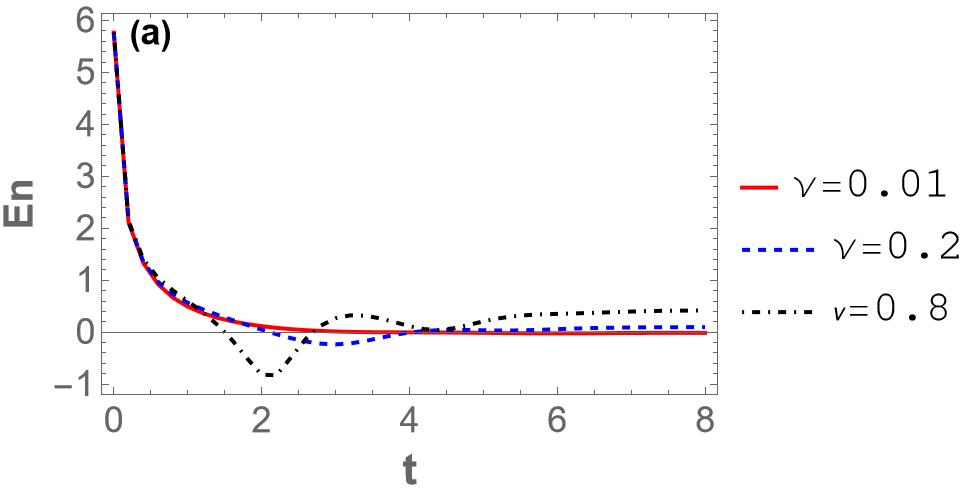}
    \hfill
    \includegraphics[width=0.45\textwidth]{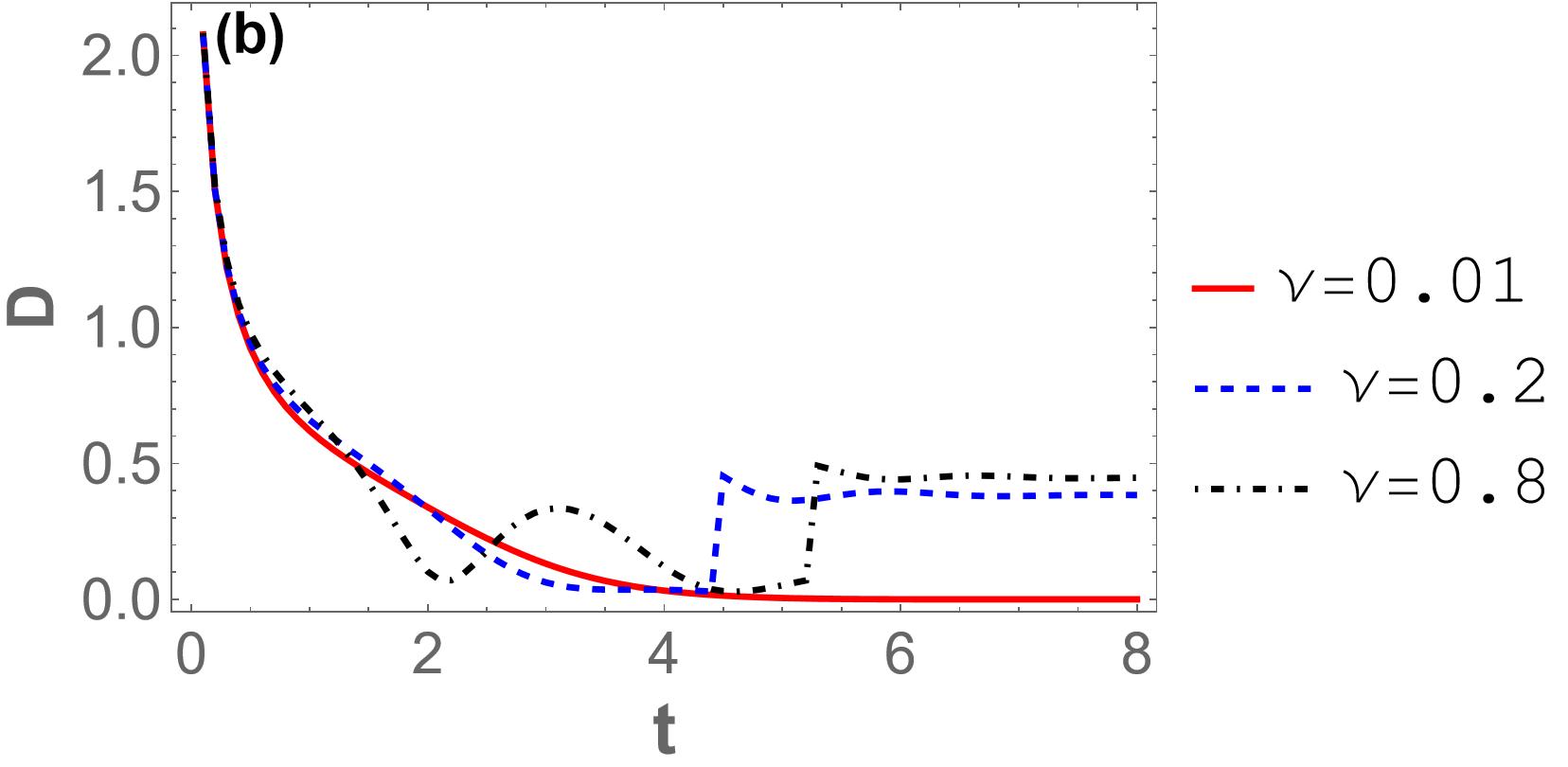}
    \vspace{0.5cm}
    \includegraphics[width=0.45\textwidth]{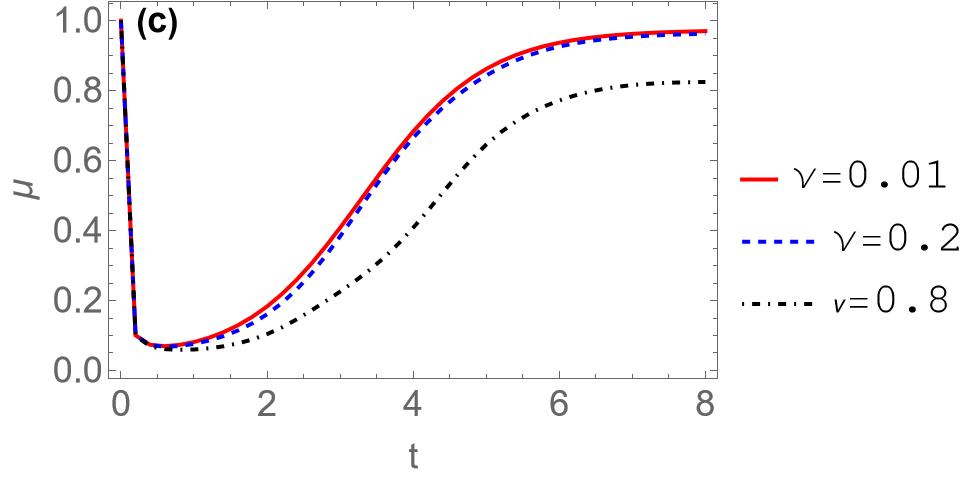}
    \caption{The evolution (a) entanglement (b) discord (c) purity for different values of coupling constant with $\varepsilon=0.5$, $r=2$, $\lambda=0.6$, $T = 0.2$ and $\omega=1$.}
    \label{fig:layout-example}
\end{figure}

Figs.~4(a)--4(c) illustrate the influence of the coupling constant on entanglement, discord and purity. 
For relatively small values of the coupling constant entanglement is in general decreasing over time, while for relatively large values of the coupling constant entanglement manifests an oscillatory behaviour. The observed oscillations in both discord and entanglement originate from the competition between the coherent energy exchange induced by the XY-type coupling term $\nu x_1 x_2$ and the continuous dissipative action of the thermal environment. This interplay produces transient revivals before the system settles into the steady state.  
In general entanglement is increasing with the coupling between the modes. We again notice that, depending on the values of the squeezing parameter, coupling constant and temperature of the bath, one can notice a sudden death of entanglement, followed by temporary revivals and suppressions of entanglement.
We can also observe in Fig.~4(a) that in the limit of asymptotically large times entanglement can survive for definite values of coupling between the modes and temperature, and its value at time infinity is increasing with the coupling constant. For comparison, we observe in Fig.~4(b) that discord has in general a non-monotonic behaviour in time, it is decreasing initially over time and tends asymptotically to a non-zero value as time approaches infinity, for a non-zero coupling between the two modes. 
The presented behaviour in time of the entanglement is the result of the competition between the influences produced by the parameters characterising the initial states of the system and of the thermal bath, and by the coupling between the modes. 
In Fig.~4(c) it is shown that purity declines more steeply for larger coupling, reflecting enhanced decoherence due to interaction with the environment. Physically, stronger dissipation $\lambda$ drives the system faster toward the thermal steady state; however, due to the specific form of the diffusion matrix and the residual XY coupling, higher $\lambda$ suppresses position-momentum fluctuations more effectively, resulting in a higher steady-state purity (lower mixedness). This counter-intuitive behaviour highlights the non-trivial role of dissipation in coupled oscillator systems.

\section{Conclusion}
In this study we have analyzed the decoherence dynamics of a system composed of two coupled asymmetric harmonic oscillators interacting with a thermal environment. By employing the Kossakowski-Lindblad master equation under the Markovian approximation, we derived the evolution equations in terms of the covariance matrix and systematically investigated the temporal behaviour of quantum correlations and purity. Starting with a squeezed vacuum state, we examined how the system and environmental parameters affect the behaviour of purity and quantum correlations, revealing distinct roles for each parameter in shaping the decoherence dynamics.

A common characteristic of the behaviours manifested by the quantum discord and entanglement is that their initial values are in general non-monotonically decreasing over time, while purity decreases initially and then stabilizes. We observed that higher temperatures lead to a faster decreasing of both quantum discord and entanglement due to enhanced thermal noise, while simultaneously reducing the steady-state purity of the system. Likewise, increasing the dissipation coefficient accelerates the decrease of quantum correlations, but leads to a higher steady-state purity, reflecting the complex influences of dissipation and coupling in the oscillator system. The squeezing parameter emerges as a crucial protective factor, with larger squeezing parameters enhancing initial quantum correlations and sustaining entanglement for longer periods.
% demonstrating more pronounced oscillatory behavior in discord evolution. 
However, increasing squeezing also accelerates the initial decoherence process, leading to a stronger decrease of the purity, by amplifying the system sensitivity to the environmental noise. At the same time, the asymmetric parameter exhibits relatively weak effects on the two types of quantum correlations and purity. 

%While it has negligible impact on quantum discord evolution and final purity, it significantly affects entanglement dynamics, with increasing asymmetry %leading to notable reductions in entanglement, particularly at higher values. This indicates the system's sensitivity to structural imbalances and suggests %that symmetric configurations are more favorable for entanglement preservation. 
%The coupling constant between the oscillators plays a dual role in the system dynamics, with stronger coupling accelerating the decrease of both %entanglement and discord, indicating more rapid information exchange with the environment and leading to steeper declines in purity.

In general, entanglement is increasing with the coupling between the modes, with the remark that for relatively small values of the coupling constant it is decreasing over time, while for relatively large values of the coupling constant entanglement manifests an oscillatory behaviour. Only in the case of uncoupled modes ($\nu=0$) entanglement is monotonically decreasing over time.  
Depending on the values of the squeezing parameter, coupling constant, dissipation and temperature of the bath, one can notice a sudden death of entanglement, possibly followed by temporary revivals and suppressions of entanglement, or sustained entanglement. In the limit of asymptotically large times entanglement can survive for definite values of coupling between the modes, temperature and dissipation, and its value at time infinity is increasing with the coupling constant. The presented behaviour in time of the entanglement is the result of the competition between the influences produced by the parameters characterising the initial states of the system and of the thermal bath, and by the coupling between the modes. For comparison, discord has in general a non-monotonic behaviour in time, however it shows a greater resilience than entanglement. Namely, discord keeps a non-zero value all the time, it tends asymptotically to a non-zero value as time approaches infinity, for a non-zero coupling between the two modes, and approaches zero asymptotically at infinite time only when the coupling is zero. The presence of coupling between the two modes induces time-dependent fluctuations in discord evolution, suggesting a complex interplay between coupling mechanism and quantum correlation preservation.
Concerning purity, it declines more steeply for larger coupling, reflecting enhanced decoherence due to interaction with the environment. 

The obtained results provide a deeper understanding of decoherence mechanisms and offer practical strategies for enhancing quantum correlation longevity. The protective role of squeezing parameters suggests viable approaches for developing robust quantum information protocols. The observed resilience of quantum discord, even beyond entanglement suppression, points to its potential utility in quantum technologies where classical correlations are insufficient. These findings establish a foundation for further investigations into non-Markovian regimes, multi-mode systems, and decoherence mitigation strategies.

\section*{Acknowledgments}

AI acknowledges the financial support received from the Romanian Ministry of Education and Research, through the Project PN 23 21 01 01/2023.

% Including bibliography
\bibliography{References}

\end{document}